\documentclass[final]{hubsarxiv}

\usepackage{listings}
\usepackage{needspace}   
\graphicspath{{figures/}}
\lstdefinestyle{python}{
  language=Python,
  basicstyle=\ttfamily\small,
  keywordstyle=\color{blue},
  commentstyle=\color{gray},
  stringstyle=\color{teal},
  breaklines=true,
  frame=single,
  numbers=left,
  numberstyle=\tiny\color{gray},
  numbersep=5pt,
}

\title{BuildOcc: A Large Language Model Occupant Agent Platform for Building Energy Research}
\shorttitle{BuildOcc: an LLM occupant agent platform}
\shortauthors{Jung}
\author{Wooyoung Jung\affmark{a,}\corrmark\orcid{0000-0003-2746-4983}}
\affil{\affmark{a}Department of Civil and Architectural Engineering and Mechanics, University of Arizona, Tucson, AZ, USA}
\corresponding{wooyoung@arizona.edu}
\paperno{2026-004}
\paperdate{September 2026}
\versionnote{v1 (2026-09-02): first release. v2 (2026-09-25): revised. v3 (2026-11-05): re-set in the HUBS Lab paper layout; content unchanged.}
\published{\emph{SoftwareX}, 2026. \url{https://doi.org/10.1016/j.softx.2026.103068}. Accepted manuscript, CC BY-NC-ND 4.0.}
\license{CC BY-NC-ND 4.0}
\abstract{Occupants are a primary source of uncertainty in building energy consumption and
management, yet existing occupant behavior models cannot fully capture adaptive, reasoned
responses that occupants form from their personal history, current context, and the type of
energy signal they receive. This study presents \textsc{BuildOcc}, an open-source
Python platform that grounds large language model-based occupant agents in the American Time Use
Survey (ATUS), a nationally representative diary dataset, using the 6,611 diary
respondents who fall within four demographic strata: employed single adults, retired couples, employed parents, and not-employed adults.
Each simulated occupant agent combines a demographic
persona drawn from ATUS population statistics, an activity scheduler that samples the day's routine from the same data, a memory stream that accumulates and
reflects on timestep-level observations, and a reasoning engine that selects each action and states the reason for it. The platform exposes a three-layer interface---Python 
library, REST application programming interface, and Model Context Protocol server---so that
any building energy tool (EnergyPlus, Home Assistant) can integrate behavioral
intelligence without bespoke coupling code. A plugin registry lets the community add new
occupant strata and alternative memory backends as separate installable
packages. Two validation tiers show that the activity scheduler reproduces its ATUS reference
distributions to within sampling noise and that demographic priors propagate into persona-consistent
agent reasoning across timesteps, establishing internal consistency across strata.
\textsc{BuildOcc} provides the building energy community with a reusable
implementation of the occupant behavioral layer, openly released at \url{https://doi.org/10.5281/zenodo.22255938} under
the Apache License~2.0 and installable via \texttt{pip install buildocc}.}

\begin{document}
\renewcommand{\labelenumii}{\arabic{enumi}.\arabic{enumii}}
\maketitle

\Needspace*{0.55\textheight}
\section*{Required Metadata}

\subsection*{Current code version}

\begin{table}[!ht]
\caption{Code metadata (mandatory)}
\label{tab:metadata}
\begin{tabular}{|l|p{5.5cm}|p{5.5cm}|}
\hline
\textbf{Nr.} & \textbf{Code metadata description} & \textbf{Please fill in this column} \\
\hline
C1 & Current code version & v1.0.1 \\
\hline
C2 & Permanent link to code/repository &
  \url{https://github.com/humanbuildingsynergy/BuildOcc} \\
\hline
C3 & Permanent link to reproducible capsule &
  \url{https://doi.org/10.5281/zenodo.22255938} \\
\hline
C4 & Legal code license & Apache License 2.0 \\
\hline
C5 & Code versioning system & Git \\
\hline
C6 & Software code languages, tools, and services used & Python 3.11+ \\
\hline
C7 & Compilation requirements, operating environments \& dependencies &
  \texttt{pip install buildocc}; requires a large language model (LLM) API key
  (Anthropic, OpenAI, Google), or local Ollama \\
\hline
C8 & Developer documentation &
  \url{https://github.com/humanbuildingsynergy/BuildOcc/blob/main/README.md} \\
\hline
C9 & Support email for questions & wooyoung@arizona.edu \\
\hline
\end{tabular}
\end{table}

\section{Motivation and Significance}
\label{sec:motivation}

Humans are complex systems that respond to environmental, social, and economic signals
in ways that are difficult to predict, introducing substantial uncertainty in building
energy use and management~\cite{hong2016advances,yan2015occupant,jung2019hitl}. However, building energy modeling often represents humans as static entities through fixed schedules (e.g., ASHRAE 90.1 reference schedules~\cite{ashrae901,ye2024loadschedules}),
which assign the same profile to every day and every occupant.
This simplification contributes to the gap between simulated and actual energy consumption~\cite{hong2016advances,yan2015occupant} and
has motivated the development of occupant behavior models that can capture the unpredictable
and adaptive nature of human decision-making~\cite{arslan2026survey}. Stochastic models, for example, improved on
fixed schedules by sampling from sensor data~\cite{jung2023occupancysim} or from national
time-use diaries: early work used first-order
Markov chains to generate occupant behaviors with realistic day-to-day
variation~\cite{richardson2008,widen2010}; later models added explicit behavior episodes with
durations~\cite{wilke2013} and higher-order transitions that distinguish occupant
types~\cite{flett2016}. Each step reproduces occupant behavior statistics more faithfully, but
all remain statistical descriptions.

Large language models (LLMs), with architectures for reasoning and memory, allow artificial intelligence (AI) agents
to simulate human-like decision-making through contextual reasoning, memory retention,
and adaptive behavior---capabilities suited to modeling occupants in buildings. Recent
studies demonstrated the potential of LLM-integrated AI agents to simulate occupant behavior
in building energy contexts, showing that they can produce more realistic and adaptive
responses to various energy signals compared to traditional
models~\cite{arslan2026survey,lyu2026llm,he2025context,qi2026energy,jung2026hema,jung2026domainknowledge}.
Specifically, Nicholas et~al.~\cite{nicholas2026generative} proposed AI agents that adopt
the memory architecture of Park et~al.'s generative agents~\cite{park2023}---combining a
persona-based core memory, a timestep-level memory stream, and periodic reflection---and
demonstrated that this representation produces qualitatively different energy demand patterns
than fixed-schedule or rule-based occupant models. Other agents extend this architecture
in different directions: conditioning behavior on indoor environmental quality~\cite{lee2026generative},
negotiating a shared setpoint between occupants~\cite{rende2025negotiating}, responding
to dynamic pricing~\cite{chen2026behavioral}, and adjusting setpoints under demand
response~\cite{deng2026llm}. However, these agents are instantiated from hand-crafted
profiles or small convenience samples~\cite{leng2025agentsense}, with few grounding their schedules in national survey data, and few being released as open platforms---leaving no shared,
survey-grounded starting point for reproducible occupant behavior research.

This study addresses that gap with \textsc{BuildOcc}, an open-source platform for
LLM-based occupant simulation. Personas and activity schedules are grounded in the
American Time Use Survey (ATUS)~\cite{bls2024atus}, and each agent carries a memory stream
that accumulates observations and periodically reflects on them, giving coherent
behavior across a simulation day. At each 15-minute timestep, a reasoning engine combines
the persona, the retrieved memories, and the current environment to select the occupant's
next action and state the reason for it. It weighs incoming demand response signals the
same way, accepting, rejecting, or deferring each with a stated reason.
A three-layer interface (Python library, REST application programming interface (API), and
Model Context Protocol (MCP) server) connects the occupant agent to EnergyPlus~\cite{energyplus2023} co-simulation,
Home Assistant, or a custom building management system, and a plugin registry lets
other groups add strata, grounding sources, and memory architectures.

\section{Software Description}
\label{sec:software}

This section first gives an overview of the platform architecture and then describes the
ATUS behavioral grounding that drives activity scheduling and persona sampling; the inputs
required at simulation initialization; the per-timestep LLM reasoning pipeline; the demand
response signal interface; the extensibility mechanisms that allow the platform to be adapted
to new research contexts; and the computational cost of running it.

\subsection{Architecture Overview}

\begin{figure}[!ht]
  \centering
  \includegraphics[width=\linewidth]{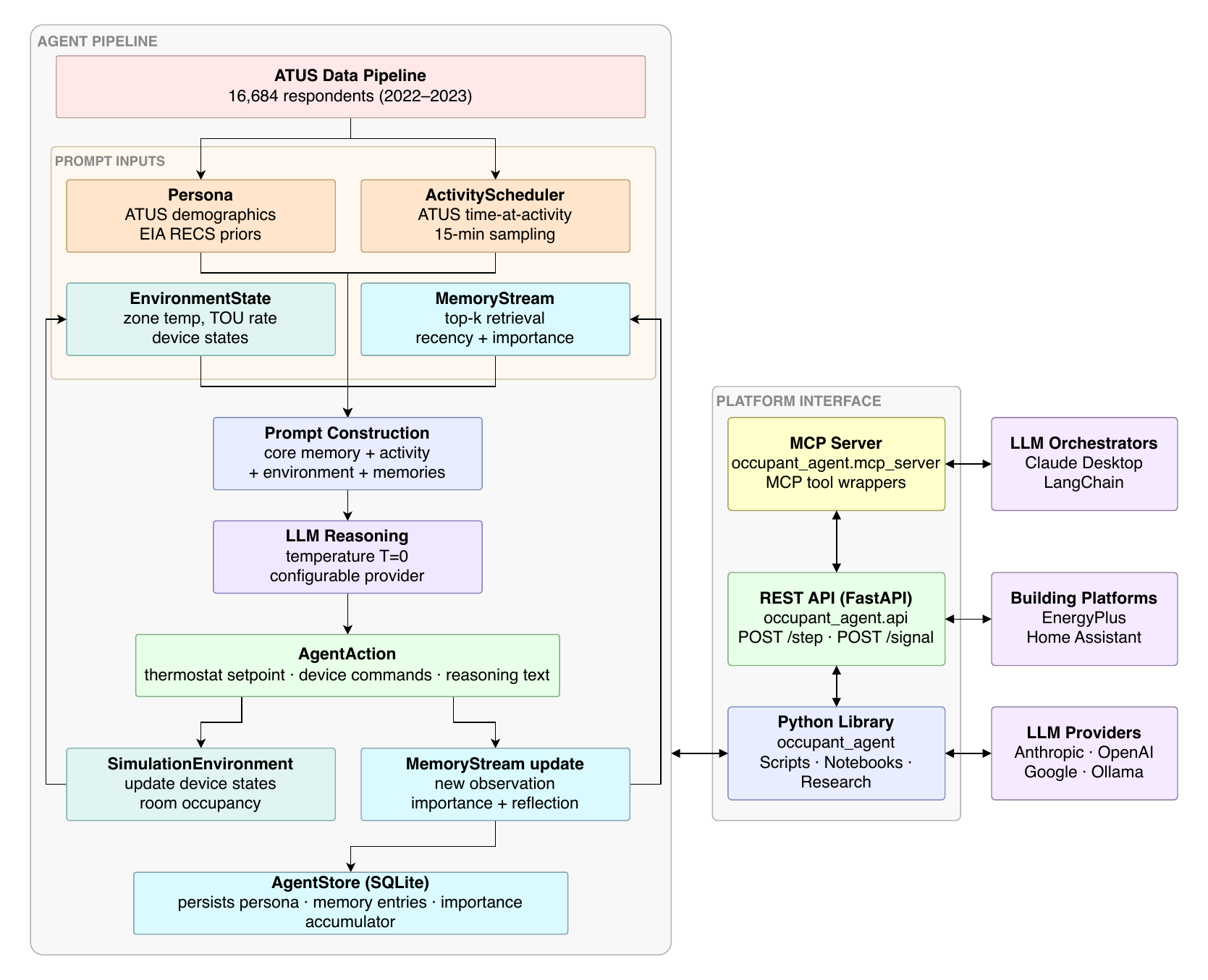}
  \caption{Architecture of \textsc{BuildOcc}. \emph{Left}: per-timestep agent
  pipeline showing how ATUS-grounded activity codes, retrieved memories, and
  environment state are assembled into an LLM prompt and how the resulting
  \texttt{AgentAction} updates both the simulation environment and the memory
  stream. \emph{Right}: three-layer platform interface through which external
  building energy tools and LLM orchestration frameworks access the agent.}
  \label{fig:architecture}
\end{figure}

\textsc{BuildOcc} is organized around two levels of architecture: an internal
\emph{agent component architecture} that defines how occupant agents perceive,
reason, and act within a simulation timestep, and a \emph{platform interface}
that exposes this functionality to external building energy tools
(Figure~\ref{fig:architecture}).

\subsubsection{Agent Component Architecture}

The \textsc{BuildOcc} occupant agent is composed of six interacting
components:

\begin{itemize}
    \item \textbf{Persona} --- a structured demographic profile assigned to one
          of four strata (O1--O4; Section~\ref{subsec:atus}), with appliance
          ownership drawn from Residential Energy Consumption Survey (RECS)~\cite{eia2023recs}
          priors, an income bracket drawn uniformly within the stratum's RECS range, and
          a comfort band set from income.
    \item \textbf{ActivityScheduler} --- samples activity at each timestep from
          stratum-, hour- and day-type-specific ATUS distributions
          (6,611 respondents; Appendix~\ref{app:strata}).
    \item \textbf{MemoryStream} --- an accumulating log of observations, actions,
          and reflections, ranked by recency and LLM-assigned importance following
          Park et al.~\cite{park2023} (Appendix~\ref{app:memory}).
    \item \textbf{LLM Reasoning Engine} --- queries a configurable provider
          (Anthropic, OpenAI, Google, or local Ollama) at temperature $T{=}0$,
          returning a structured action with a natural-language reasoning string.
    \item \textbf{SimulationEnvironment} --- maintains device states, room occupancy,
          and thermostat setpoint between timesteps, exposing the current state before
          each decision and applying the returned action after it.
    \item \textbf{AgentStore} --- persists the agent's cognitive state (persona,
          memories, importance accumulator) to SQLite after each operation. Environment
          state is not stored: the calling program sends it with every request, so
          the server keeps nothing between calls.
\end{itemize}

\subsubsection{Platform Interface}
\label{subsec:interface}

\textsc{BuildOcc} is distributed as the package \texttt{buildocc} and exposes the agent through three integration
layers:

\begin{enumerate}
    \item \textbf{Python library} (\texttt{occupant\_\allowbreak agent}) --- direct access to all agent components; the recommended path for batch
           simulations.
    \item \textbf{REST API} (\texttt{occupant\_agent.api}) --- a web server (FastAPI~\cite{fastapi2023}) that keeps no
          state between requests, exposing agent initialization, per-timestep
          decisions (\texttt{POST /step}), signal delivery (\texttt{POST /signal}), and
          state query (\texttt{GET /state}); suitable for EnergyPlus co-simulation
          callbacks and any platform that can issue ordinary web (HTTP) requests.
    \item \textbf{MCP server} (\texttt{occupant\_agent.mcp\_server}) --- wraps the REST
          API as MCP~\cite{mcp2024} tools, so any MCP-compatible host
          (Claude Desktop, LangChain, Home Assistant) can drive a simulation without
          additional integration code.
\end{enumerate}

\subsection{ATUS Behavioral Grounding}
\label{subsec:atus}

Activity schedules are derived from ATUS 2022--2023 microdata.
Time-at-activity probabilities are aggregated across nine categories
(sleeping, work, food preparation, laundry, television viewing, eating,
exercise, travel, and other) by hour of day, stratified by demographic group and day type (weekday / weekend).
The method follows the ``what is the person doing at this moment?'' framing:
for each diary episode $[t_\text{start}, t_\text{stop}]$, the method queries whether the
episode covers the half-hour mark of each hour $h$, then computes a weighted probability across respondents using the ATUS final
person weight (\texttt{TUFINLWGT}). Weighting makes each probability an
estimate for the U.S. population. The formal computation
is given in Appendix~\ref{app:atus}. The scheduler draws each timestep
independently from the hourly distribution, so it reproduces the hourly marginals
without modeling how long an activity lasts, and episode lengths are consequently
shorter than the diaries record (Section~\ref{subsec:tier1}); no duration model is
implemented in this release.

Four demographic strata are defined (Table~\ref{tab:strata}), selected to span
the variation in building occupancy and demand response participation that is most
relevant to residential energy research (Appendix~\ref{app:strata}). Of the
16,684 respondents surveyed, 6,611 fall within these four strata.

\begin{table}[!ht]
\caption{Demographic strata}
\label{tab:strata}
\begin{tabular}{llp{0.55\linewidth}}
\toprule
\textbf{ID} & \textbf{Label} & \textbf{ATUS filter} \\
\midrule
O1 & Employed single adult & Full-time employed, lives alone, age 25--44 \\
O2 & Retired couple        & Age 65+, not employed, spouse or partner present \\
O3 & Employed parent       & Full-time employed, children present, age 35--54 \\
O4 & Not-employed adult    & Not employed (unemployed or not in the labor force),
     age 25--44 \\
\bottomrule
\end{tabular}
\end{table}

Work-from-home (WFH) behavior is handled separately: a per-agent WFH
probability encodes the fraction of working days spent at home (the share of each
stratum's work time that ATUS respondents report at home; Table~\ref{tab:strata_rationale}),
and a per-day Bernoulli draw determines WFH
status for each simulated day. The WFH flag is included in the agent's
per-timestep prompt so that the LLM reflects both the ATUS activity code and the
work location, preventing the scheduling model from assigning ``work'' activities
that place the agent outside the home on days it is working remotely.

Each stratum uses stratum-specific appliance ownership priors from the U.S. Energy Information Administration's
RECS~\cite{eia2023recs,jung2026thermostat}
and a comfort band (in $^\circ$C) that widens for lower income brackets,
reflecting the documented tendency of lower-income households to accept wider thermal deviations to reduce heating, ventilation, and air-conditioning (HVAC)
costs~\cite{gyamfi2011price,eia2023recs,kim2026energyburden,jung2026pcs}.

\subsection{User-Provided Inputs}

Table~\ref{tab:inputs} lists all inputs required at initialization; the only
per-timestep value the calling application must supply is zone temperature,
since it depends on the building's physical response to occupant actions. To lower the
barrier to first use, the library ships preset helpers for a representative household
device inventory and room layout. A command-line interface offers the same presets and a plain-text (YAML) configuration file for users who
prefer not to write Python.

\begin{table}[!ht]
\caption{User-provided inputs to \textsc{BuildOcc}. All are specified once
         at initialization, except zone temperature, which the calling
         platform supplies at each timestep.}
\label{tab:inputs}
\begin{tabular}{llp{0.55\linewidth}}
\toprule
\textbf{Input} & \textbf{Type} & \textbf{Description} \\
\midrule
Stratum            & String  & Demographic profile: O1--O4 \\
Seed               & Integer & Controls persona attribute sampling, activity scheduling, and work-from-home day draws; the same seed produces an identical agent across runs. \\
LLM provider       & String  & Anthropic, OpenAI, Google, or Ollama \\
Device inventory   & List    & Appliance IDs, initial on/off states, rated power (W) \\
Room list          & List    & Rooms the occupant can occupy \\
Thermostat setpoint & Float  & Initial comfort setpoint (°C) \\
Zone temperature   & Source  & Fixed value, CSV file, or live building platform \\
Outdoor temperature & Source & Fixed value, CSV file, EnergyPlus weather (EPW) file, or callable \\
Time-of-use (TOU) rate schedule & Source & Fixed value, CSV file, or tariff function \\
\bottomrule
\end{tabular}
\end{table}

Without a live link, zone temperature can be a constant or a pre-computed
comma-separated values (CSV) file; a one-week EnergyPlus sample is included.

\subsection{LLM Reasoning Pipeline}

At each 15-minute timestep the agent resolves its current ATUS code to a
natural-language activity and an occupancy flag (a sleeping or laundry activity moves
the agent to the matching room by rule, without a model call), retrieves up to five
memories ranked by
recency and importance (Appendix~\ref{app:memory}), and assembles
these with the persona and the environment state into a system and user prompt. The
model returns one of four action types (Table~\ref{tab:action_types}) as a
structured-text (JSON) object
carrying a target, a value, and a one-sentence memory note, which is appended to the
stream with an LLM-assigned importance score. When the cumulative importance exceeds a
threshold (default 100), a reflection pass distills recent observations into three
higher-order insights. Figure~\ref{fig:architecture} shows the same sequence.

\begin{table}[!ht]
\caption{The four action types returned by \texttt{step()} at each timestep,
         with their decision trigger and a representative example.
         Demand response is handled separately via \texttt{receive\_signal()}
         (Section~\ref{subsec:signals}).}
\label{tab:action_types}
\centering
\small
\begin{tabular}{@{}lp{0.34\linewidth}p{0.34\linewidth}@{}}
\toprule
\textbf{Action type} & \textbf{Trigger} & \textbf{Example} \\
\midrule
\texttt{do\_nothing}         & No environmental or activity condition warrants intervention      & Agent is sleeping; no device is running \\
\texttt{adjust\_thermostat}  & Zone temperature leaves the comfort band, or TOU rate context & Raises setpoint during peak window to reduce HVAC cost \\
\texttt{toggle\_device}      & Activity--appliance alignment (on) or inactivity (off)           & Turns off lights when leaving for work \\
\texttt{move\_room}          & Activity context suggests transitioning to a different room       & Moves to bedroom at 23:00 for sleeping \\
\bottomrule
\end{tabular}
\end{table}

The engine samples at $T{=}0$ by default, so the model returns its single most
probable output and a fixed provider, model version, and prompt reproduce the same
action on every run. That determinism is
relative to one set of model weights: it does not hold across providers or across
successive versions of a hosted model. Studies that must stay reproducible over time can run a fixed, locally hosted open
model, whose weights do not change.

\subsection{Demand Response Signal Interface}
\label{subsec:signals}

Alongside the per-timestep pipeline, \textsc{BuildOcc} provides a dedicated
signal delivery interface that can be invoked at any point in a simulation.
Three signal types are supported (Table~\ref{tab:signal_types}), following
established behavioral intervention
typologies~\cite{gyamfi2011price,albadi2008summary,jung2026ecofeedback}. Signals are delivered
by passing a signal type, a free-text message, and the current environment
state to the agent, which returns a response code (\textit{accepted},
\textit{rejected}, or \textit{deferred}) alongside a natural-language
reasoning string drawn from the agent's memory context. Keeping the three
types mutually exclusive enables factorial experimental designs in which
signal type and demographic stratum are the primary independent variables.

\begin{table}[!ht]
\caption{Demand response signal types supported by \textsc{BuildOcc},
         classified by behavioral lever and illustrated with representative
         example messages. All three types share the same delivery mechanism
         and response codes (\textit{accepted}, \textit{rejected},
         \textit{deferred}).}
\label{tab:signal_types}
\centering
\small
\begin{tabular}{@{}lp{0.2\linewidth}p{0.47\linewidth}@{}}
\toprule
\textbf{Type} & \textbf{Behavioral lever} & \textbf{Example message} \\
\midrule
A — Direct command  & Regulatory compliance &
  ``Turn off your HVAC for 30 min to support grid stability
    during today's peak demand event.'' \\[4pt]
B — Educational     & Economic information  &
  ``Your HVAC costs 3 times more per kWh at the current peak rate ---
    raising the setpoint 1$^\circ$C saves an estimated \$0.35 today.'' \\[4pt]
C — Social norm     & Peer comparison       &
  ``75\% of similar households in your area have reduced HVAC use
    during today's peak event.'' \\
\bottomrule
\end{tabular}
\end{table}

\subsection{Platform Extensibility}

\textsc{BuildOcc} is designed for adaptation across research contexts without
modifying the core library. Extensibility is organized along three independent
axes:

\begin{itemize}
    \item \textbf{Occupant population} (\texttt{BasePersona}): new demographic
          profiles --- commercial occupants, hotel guests, non-US populations,
          or any custom stratum --- are registered by implementing
          \texttt{BasePersona}.
    \item \textbf{Activity grounding} (\texttt{BaseScheduler}): the built-in
          scheduler uses ATUS microdata; any other grounding source (a Markov chain,
          a physics-based occupancy model, or alternative activity data) can be
          substituted by implementing \texttt{BaseScheduler}.
    \item \textbf{Memory architecture} (\texttt{BaseMemoryStream}): the
          built-in recency-plus-importance retrieval can be replaced with
          retrieval by semantic similarity (embeddings) or any other scheme by subclassing
          \texttt{BaseMemoryStream}.
\end{itemize}

Extensions declare themselves in their own package configuration (Listing~1) and are
found automatically when the library loads, so a third party can add a stratum or
scheduler without modifying \textsc{BuildOcc}'s own code:

\begin{lstlisting}[style=python, caption={pyproject.toml entry-point declarations for a third-party stratum and scheduler.}]
[project.entry-points."occupant_agent.strata"]
P5 = "my_package.personas:LowIncomeElderlyAlone"

[project.entry-points."occupant_agent.schedulers"]
homer = "my_package.schedulers:HomerScheduler"
\end{lstlisting}

Transfer to another country is a data-substitution task rather than a
redesign: the scheduler consumes a table of activity probabilities by stratum, hour,
and day type, so any national time-use survey following harmonized diary conventions~\cite{osman2021}---the Multinational Time Use Study~\cite{fisher2013mtus} or the Harmonised European Time Use Surveys (HETUS)~\cite{eurostat2019hetus}, for example---can be substituted by
regenerating that table behind \texttt{BaseScheduler}.
What does not transfer automatically is the persona layer, whose appliance-ownership
and income priors come from a US survey (RECS) and would need a national equivalent.

Additional context---comfort indices~\cite{fanger1970thermal}, language-based comfort
and heat-strain signals~\cite{jung2026langsensor,babonayeng2026heatstrain}, or air
quality readings---can be injected by extending the agent class or by passing free text
at each timestep, through either interface. The environment and signal schemas are
fixed for reproducibility, and a testing module lets extension authors check, without an API key, that a new
stratum or scheduler meets the platform's contract.

\subsection{Computational Cost and Scalability}

Because the agent queries a language model at most timesteps, cost and runtime scale
linearly with simulated agent-days. A 15-minute timestep gives at most 96 calls per
simulated day per agent, plus occasional reflection calls, with responses capped at
512 tokens. Measured with Claude Haiku 4.5 for per-timestep decisions and Claude Sonnet 4.6 for
reflection, over two simulated agent-days for each of the four strata
(September 2026 list prices), one agent-day required 85.5--90.0 per-timestep calls and
1.5 reflection calls, consumed 111,000--122,000 prompt and 13,700--15,300
completion tokens, and cost \$0.19--\$0.21. Fewer than 96 calls are issued because room transitions that follow directly from
the sampled activity, such as moving to the bedroom when sleep begins, are resolved
without a model call; they are still counted as moves in Table~\ref{tab:actions}. Wall-clock time was 2.1--4.3\,s per timestep and is dominated by provider
latency, not local computation. Two mechanisms reduce this cost.
Any local open-weight model served through Ollama removes per-call cost
entirely, at the price of reduced reasoning quality, and scheduler-only
studies---including the Tier~1 validation reported in Section~\ref{subsec:tier1}---require no model
calls at all.

These figures are per agent. Agents are independent---each carries its own persona,
memory, and schedule, and the REST API is stateless---so a multi-occupant household
runs one agent per occupant, with call volume scaling linearly and the agents running in parallel.

\section{Illustrative Examples}
\label{sec:example}

This section demonstrates the platform in three parts: a walkthrough of a single
agent through a demand response event, followed by two quantitative tiers. Tier~1
validates the \textit{ActivityScheduler}, testing whether ATUS-grounded sampling
produces category distributions matching empirical ATUS reference data relative to a
deterministic rule-based baseline; Tier~2 validates the \textit{LLM agent decisions},
testing whether each stratum's demographic priors propagate into distinguishable
behavioral patterns under identical environment conditions. All LLM results in this section
use Anthropic models at $T{=}0$ with a 512-token response cap:
\texttt{claude-haiku-4-5-20251001} for per-timestep decisions and
\texttt{claude-sonnet-4-6} for reflection synthesis; Tier~1 involves no model calls.

\subsection{Demand Response Walkthrough}

To illustrate agent behavior during a residential demand response event, this subsection
traces an O1 agent (28-year-old employed male with an annual family income of
\$30,000--\$34,999, comfort band $\pm$1.1$^\circ$C around the setpoint) through the peak electricity period of a summer weekday
evening (18:00--19:45, TOU rate \$0.22/kWh). Zone temperatures come from pre-computed EnergyPlus output included in the
repository~\cite{doe2023prototypes}; outdoor temperature follows a synthetic
summer-day profile. The simulation loop is:

\begin{lstlisting}[style=python, caption={Running the BuildOcc simulation loop with
      the sample zone temperature CSV.}]
# Full imports available in examples/simulation_loop.py (repository)
from datetime import datetime, timedelta

agent     = OccupantAgent.from_stratum("O1", seed=42)
scheduler = ActivityScheduler(stratum="O1", seed=42)
zone_temp = zone_temp_from_csv(...)   # EnergyPlus zone temperatures
price_message = ("Your HVAC is running during peak hours (4-9pm) at $0.22/kWh - "
                 "three times the off-peak rate. Raising the setpoint by 1 C saves ~$0.35 today.")

sim = SimulationEnvironment(
    initial_devices=persona_devices(agent.persona),
    initial_rooms=[RoomState(room_id="living_room", occupied=True)],
    thermostat_setpoint=22.0,
    outdoor_temp_fn=summer_day_temp,
    tou_rate_fn=peak_tou_rate,
)

base = datetime(2025, 8, 11, 18, 0)
for i in range(8):
    ts     = base + timedelta(minutes=15 * i)
    env    = sim.observe(ts, zone_temp(ts))
    if i == 2:
        agent.receive_signal("B", price_message, env)
    action = agent.step(env, atus_code=scheduler.sample(ts))
    sim.apply(action, ts)
    print(ts.strftime("%H:%M"), action.action_type, action.reasoning)
\end{lstlisting}

At 18:00 (Step~1), the zone is 23.9$^\circ$C against a 22$^\circ$C setpoint, with
35$^\circ$C outdoors and the peak rate in force. The sampled activity is work-related
travel, so the agent is on its way home and raises the setpoint to 23.5$^\circ$C:

\begin{quote}
\textit{``I'm away from home during peak rate hours (\$0.22/kWh), so raising the
setpoint to 23.5$^\circ$C will reduce HVAC runtime and save money while I'm not there
to notice the slight warmth. The current 1.9$^\circ$C deviation is already outside my
comfort band, so this adjustment is justified both for cost and to prevent further
temperature drift.''}
\end{quote}

At 18:15 the agent arrives and switches on the television. At 18:30 (Step~3) a
Type~B educational signal is delivered: \textit{``Your HVAC is running during peak hours (4--9pm) at
\$0.22/kWh---three times the off-peak rate. Raising the setpoint by 1$^\circ$C saves
\textasciitilde\$0.35 today.''} The agent declines,
on the strength of its own prior action:

\begin{quote}
\textit{``I'm comfortable right now at 23.9$^\circ$C and just settled in to watch TV.
Raising the setpoint by 1$^\circ$C would push me toward discomfort (my limit is
$\pm$1.1$^\circ$C from setpoint), and \$0.35 savings isn't worth feeling warm and
uncomfortable for the next few hours.''}
\end{quote}

For the next hour the agent holds the setpoint, recalling at each step that it
adjusted the thermostat a short time ago; at 19:30, when the sampled activity turns to
sleep, it switches the television off. This cross-timestep coherence---recalling an
earlier decision and judging a new request against it---is what a fixed schedule
cannot represent. The zone temperature comes from the pre-computed file and does not
respond to the setpoint change, so the walkthrough is open-loop; closing that loop is
the role of the co-simulation interface (Section~\ref{subsec:interface}).


\subsection{Tier 1: Sampler Correctness}
\label{subsec:tier1}

\begin{figure}[!ht]
  \centering
  \includegraphics[width=0.82\linewidth]{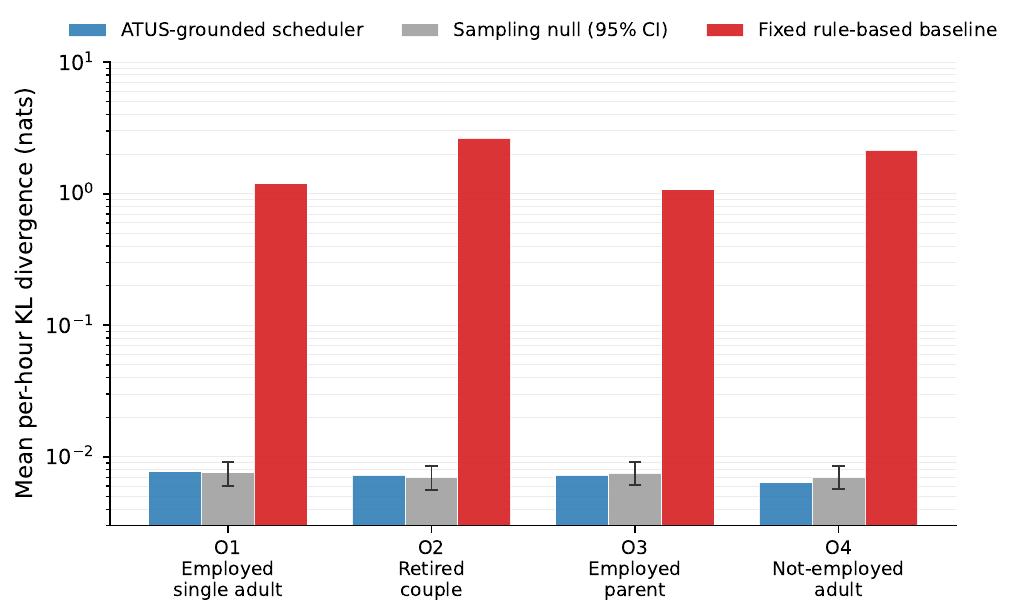}
  \caption{Mean per-hour KL divergence (nats, log scale) between simulated and
  ATUS empirical activity distributions over 180 simulated days, weekdays. The
  ATUS-grounded conditions sit at the sampling-noise floor for this sample size
  (Table~\ref{tab:kl}).}
  \label{fig:kl}
\end{figure}

The ATUS-grounded scheduler is compared with a deterministic rule-based baseline by
Kullback--Leibler (KL) divergence, which measures how far the simulated distribution
departs from the reference. For each stratum, 180 days of 96 timesteps (17,280 observations) are
simulated and the hourly category distribution is scored against the ATUS reference
using:

\begin{equation}
  D_\text{KL}(P \| Q) = \sum_{c} P(c) \log \frac{P(c)}{Q(c) + \varepsilon}
  \label{eq:kl}
\end{equation}

where $P$ is the ATUS reference distribution, $Q$ is the simulated distribution,
and $\varepsilon = 10^{-9}$ is a smoothing constant. Table~\ref{tab:kl} and
Figure~\ref{fig:kl} report the mean per-hour KL divergence over the weekday
distribution, which places 520 simulated timesteps in each hourly cell. The scheduler
samples from the same table it is scored against, so the benchmark for a correct
sampler is not zero: at finite $n$ the realized distribution departs from $P$ by
chance alone. Table~\ref{tab:kl} therefore also reports a null obtained by drawing 520
samples per hour directly from $P$ and scoring them with Equation~\ref{eq:kl}, over
1,000 repetitions.

\begin{table}[!ht]
\caption{Tier 1 sampler correctness. Mean per-hour KL divergence (nats, natural-log units) against the
         ATUS reference, over 180 simulated days of 96 timesteps (520 weekday
         observations per hourly cell). The null is the divergence expected from a
         correct sampler at the same $n$, with its 95\% confidence interval (CI), drawn
         directly from the reference
         distribution over 1,000 repetitions. Lower is better.}
\label{tab:kl}
\centering
\begin{tabular}{lrrr}
\toprule
\textbf{Stratum} & \textbf{ATUS} & \textbf{Sampling null} & \textbf{Fixed} \\
                 & \textbf{scheduler} & \textbf{(95\% CI)} & \textbf{baseline} \\
\midrule
O1 (Employed single) & 0.0077 & 0.0076 (0.0060--0.0091) & 1.19 \\
O2 (Retired couple)  & 0.0072 & 0.0070 (0.0056--0.0085) & 2.64 \\
O3 (Employed parent) & 0.0073 & 0.0075 (0.0061--0.0091) & 1.07 \\
O4 (Not-employed)    & 0.0064 & 0.0070 (0.0057--0.0085) & 2.13 \\
\bottomrule
\end{tabular}
\end{table}

\begin{figure}[!ht]
  \centering
  \includegraphics[width=\linewidth]{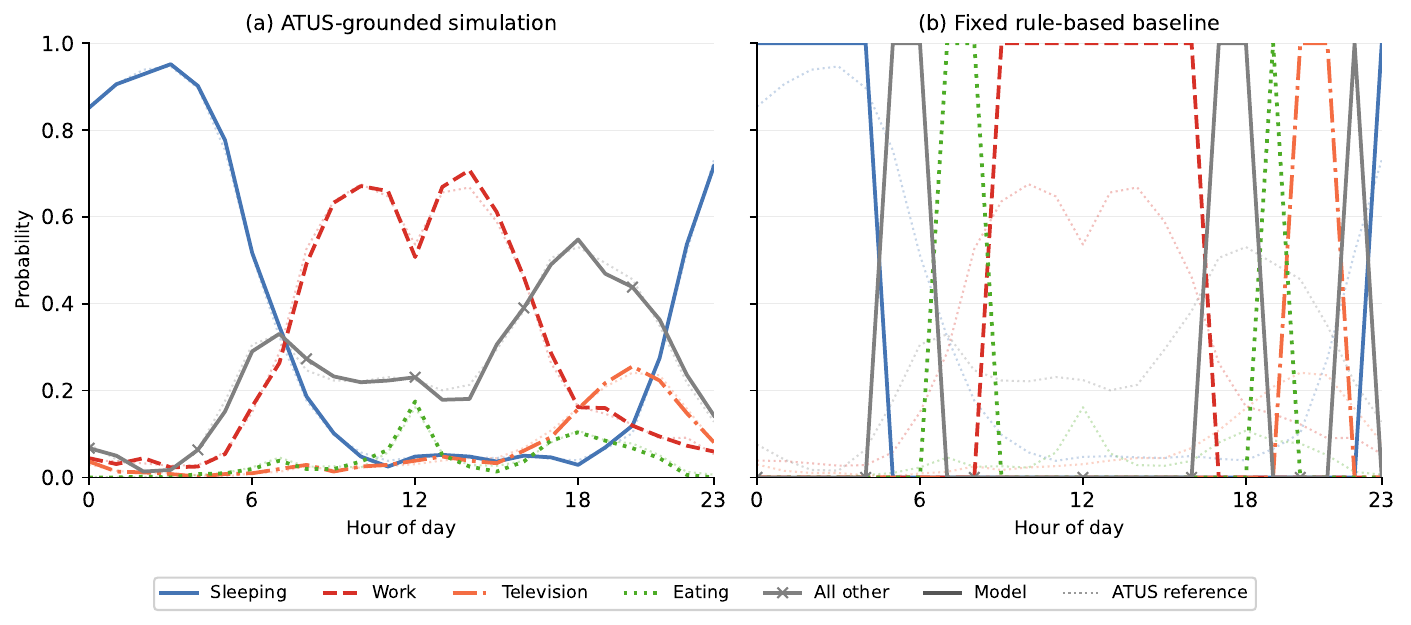}
  \caption{Hourly activity probability distributions for stratum O1 (employed single
  adult, weekday). Four load-relevant categories are drawn individually (sleeping,
  work, television, and eating) and the remaining five are pooled as ``all other''.
  Curves are population probabilities over the weekdays of 180 simulated days (520
  timesteps per hour, as in Table~\ref{tab:kl}). (a) ATUS-grounded
  simulation: bold lines track the dotted reference. (b) Fixed rule-based baseline:
  bold lines depart from the reference and misassign the O1 late-evening peak.}
  \label{fig:activity}
\end{figure}

Every ATUS-grounded condition falls inside or below its sampling null, so the
scheduler reproduces its reference tables to within sampling noise. The fixed
baseline, which assigns deterministic activities by hour-of-day rules (e.g., sleeping
23:00--06:59, work 09:00--16:59) without demographic conditioning, diverges two orders
of magnitude more, most of all for the retired (O2) and not-employed (O4) strata, whose
schedules depart furthest from the employed-adult assumption.
Figure~\ref{fig:activity} shows the contrast category by category for stratum O1 on
weekdays: the ATUS-grounded simulation (panel a) closely tracks the reference
distributions, while the baseline (panel b) collapses daytime activity into sharp step
functions and misses the gradual evening wind-down.

Per-hour marginals do not constrain episode length, however. Because each timestep is
drawn independently, episode lengths are geometrically distributed by construction, each step ending the
episode with a fixed probability: simulated sleep
episodes average 52--71~min across strata against 314--338~min in the diaries, and
agents make 49--54 activity transitions per simulated day.

\subsection{Tier 2: Persona-Driven Behavioral Differentiation}

Tier~2 tests whether each stratum's demographic priors produce distinguishable
behavior. O1--O4 were each run for three seeds over one weekday (96~timesteps) under
identical conditions---the same zone temperature, device configuration, and
time-of-use tariff (\$0.22/kWh peak, \$0.08/kWh off-peak)---so that any difference is
attributable to the persona.

\begin{figure}[!ht]
  \centering
  \includegraphics[width=\linewidth]{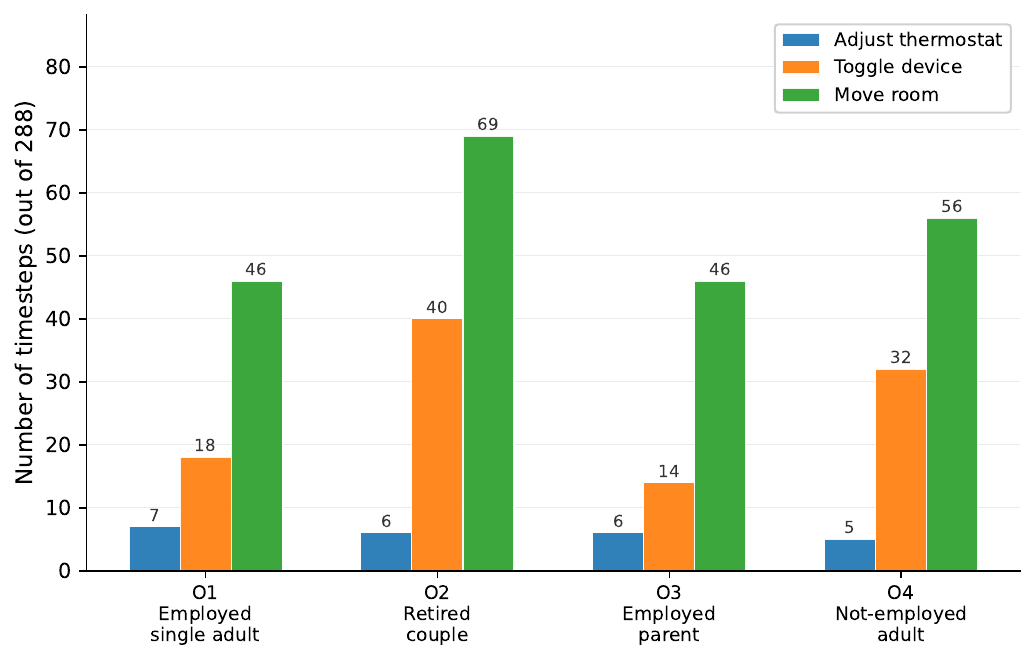}
  \caption{Active action counts per stratum across 288 timesteps (3 seeds
  $\times$ 96 steps, identical environment conditions). Do-nothing accounts
  for 60--77\% of all timesteps across strata and is excluded to isolate
  cross-stratum differences in active decision-making. Bar labels show
  absolute counts.}
  \label{fig:strata_actions}
\end{figure}

Table~\ref{tab:actions} reports action-type distributions and mean thermostat
setpoints during the TOU peak window (4--9~pm), and Figure~\ref{fig:strata_actions}
visualizes the active-action counts. The ordering follows daytime home occupancy: O2,
at home all day, is the most active stratum, with the fewest inactive timesteps and
the most device toggling and room movement, followed by O4; the two employed strata
are the least active. Of the moves, 17--25 per stratum follow by rule from a sampled
sleeping or laundry activity; the remaining 26--44 are model decisions. Every stratum
ends the peak window with its setpoint 1.2--1.6$^\circ$C above the 22$^\circ$C it
started at, O2 highest. Thermostat adjustments are rare in every stratum (1.7--2.4\% of timesteps),
so the strata differ in what they do with devices and rooms, not in how often they
touch the thermostat.

\begin{table}[!ht]
\caption{Action-type distribution (\%) per stratum: mean over three seeds,
one weekday each (96 steps), giving 288 timesteps per stratum. At this sample size
one percentage point is 2.9 timesteps, so differences of a few tenths should not be read as meaningful. Peak SP is the mean thermostat setpoint during
TOU peak hours (4--9~pm, \$0.22/kWh).}
\label{tab:actions}
\footnotesize
\begin{tabular}{lrrrrr}
\toprule
\textbf{Stratum} & \textbf{Inactive} & \textbf{Thermostat} & \textbf{Toggle} &
\textbf{Move} & \textbf{Peak SP (\textdegree{}C)} \\
\midrule
O1 (Employed single) & 75.3 & 2.4 &  6.2 & 16.0 & 23.2 \\
O2 (Retired couple)  & 60.1 & 2.1 & 13.9 & 24.0 & 23.6 \\
O3 (Employed parent) & 77.1 & 2.1 &  4.9 & 16.0 & 23.4 \\
O4 (Not-employed)    & 67.7 & 1.7 & 11.1 & 19.4 & 23.2 \\
\bottomrule
\end{tabular}
\end{table}

To assess demand response participation, three standardized signals were delivered
to each stratum (five seeds each) after a 10-timestep warm-up from 16:00 at peak rates that seeds each agent's memory: Type~A (direct
command: ``Turn off your HVAC for 30~minutes''),
Type~B (educational price signal: a cost comparison stating the peak rate as three
times off-peak),
and Type~C (social norm: ``75\% of similar households have reduced HVAC use'').
Table~\ref{tab:signals} reports the number of accepting agents per stratum and
signal type. Direct commands (Type~A) were declined in all but one case: the agents
reasoned that a 30-minute HVAC shutdown on a 35$^\circ$C afternoon would carry the
zone past their comfort band. Educational price signals (Type~B) were accepted in
every stratum, from one agent in five (O3) to three in five (O4), and every
acceptance cited the concrete dollar figure together with headroom inside the comfort
band. Social norm signals (Type~C) drew no acceptance in any stratum. Field evidence on the
nearest interventions is split: appeals to conserve during repeated peak events lose
their effect by the third event~\cite{ito2018moral}, whereas periodic reports
comparing a household's use with its neighbors' sustain savings of about
2\%~\cite{allcott2011social}. Type~C is a single peak-event message, so only the
first finding bears on it.

\begin{table}[!ht]
\caption{Signal compliance per stratum and signal type, reported as the number of
accepting agents out of five random seeds. Type~A: direct command; Type~B: educational
price signal; Type~C: social norm nudge.}
\label{tab:signals}
\begin{tabular}{l rrr}
\toprule
\textbf{Stratum} & \textbf{Type A} & \textbf{Type B} & \textbf{Type C} \\
\midrule
O1 (Employed single) & 0/5 & 2/5 & 0/5 \\
O2 (Retired couple)  & 0/5 & 2/5 & 0/5 \\
O3 (Employed parent) & 0/5 & 1/5 & 0/5 \\
O4 (Not-employed)    & 1/5 & 3/5 & 0/5 \\
\bottomrule
\end{tabular}
\end{table}

These results show that the ATUS demographic priors propagate into
distinct LLM-driven behavioral patterns: daytime home occupancy drives device
toggling and room movement, price information is the only signal accepted in every
stratum, and neither a direct command nor a social comparison moves an agent that is
already inside its comfort band. Three constraints
bound the interpretation. First, Tier~2 establishes
\emph{internal consistency}: agents instantiated from different demographic priors
behave differently, and in directions consistent with those priors. It does not
establish behavioral realism, which would require comparison against measured
occupant data and is left to future work~\cite{jung2026simusers}. Second, each stratum is run for one weekday per seed (three seeds for action
distributions, five for signals), which shows direction but cannot establish
significance; multi-day designs were constrained by API cost. Third, a single
environment configuration is used, which limits generalizability across climate zones
and occupancy contexts.

\section{Impact}
\label{sec:impact}

\textsc{BuildOcc} lowers the barrier to four lines of research that are difficult
to pursue with fixed-schedule or stochastic occupant models:

\textbf{Survey-grounded behavioral agents.} \textsc{BuildOcc} packages
population-calibrated occupant agents as a reusable component rather than a
per-project infrastructure investment. Research groups can introduce
demographically differentiated occupant behavior into any building
energy study without constructing a grounding pipeline from scratch.

\textbf{Demographic behavioral experiments.} \textsc{BuildOcc}'s combination of
ATUS-grounded demographic strata, persona income brackets, and configurable
intervention signals enables controlled factorial studies of occupant behavior.
Each agent reasons from its demographic persona and
accumulated memory at every timestep, producing adaptive responses that reflect
personal history and current context rather than predetermined patterns. Because
both the activity schedule and the agent's reasoning are conditioned on the same
demographic persona, researchers can hold environment and intervention constant
while varying stratum --- or vary the intervention while holding stratum constant
--- isolating the contribution of socioeconomic and life-stage factors to any
behavioral outcome. The Tier~2 signal experiment (Table~\ref{tab:signals})
illustrates this: educational price signals (Type~B) were accepted in every stratum,
from one agent in five (O3) to three in five (O4), while direct commands were
declined in all but one case and social comparison in every case --- differences in
direction that a fixed schedule cannot produce at all, having no mechanism to respond
to a signal. The signal types
correspond to established intervention typologies in behavioral energy
research~\cite{gyamfi2011price,albadi2008summary}, so that results can be set beside
field studies that use the same typology.

\textbf{Platform-neutral integration.} The MCP server layer means that any tool
with MCP support---including Claude Desktop, LangChain, energy scheduling
systems~\cite{elmakroum2026agentic}, and custom orchestration
frameworks---can drive the occupant agent inside a building simulation without
writing integration code. The REST API provides an equivalent path for
EnergyPlus co-simulation and Home Assistant integrations~\cite{lyu2026llm,jung2026hema}.

\textbf{Community extensibility.} The plugin registry and abstract base classes
lower the barrier for researchers who want to add custom demographic strata,
alternative behavioral data sources or activity grounding pipelines, or novel memory
architectures. The conformance test suite enables rapid validation of extensions before
publication.

\section{Conclusions}
\label{sec:conclusions}

\textsc{BuildOcc} delivers a survey-grounded, three-layer occupant agent platform
that the building energy community can use, extend, and build upon without
reconstructing the behavioral layer for each study. Tier~1 confirms that the scheduler
reproduces the ATUS activity distributions to within sampling noise; Tier~2 shows that
demographic priors propagate into differentiated agent reasoning, though it does not
by itself establish that the resulting behavior is realistic.

Beyond the implementation, the platform treats occupant behavior as a structured,
demographically parameterizable experimental variable rather than an uncontrolled
source of uncertainty. The ATUS-derived strata, frozen schema,
and plugin registry are intended to make that variable reproducible and extendable
across independent research groups.

Current limitations include a single-action constraint per timestep; ATUS-grounded
strata covering US demographic profiles only; activity grounding restricted to
primary activities, since ATUS records only one activity at a time; and a
memory-importance score that the agent assigns to itself and then reuses for
retrieval, without external calibration or a feedback path that would correct a
mis-scored entry. Subsequent phases
extend the platform with multi-agent household simulation, non-US occupancy data,
and external behavioral benchmarks including ecobee Smart Thermostat
field data~\cite{doe2021ecobee} matched to the four ATUS strata.

\renewcommand{\thetable}{\arabic{table}}
\setcounter{table}{8}
\section*{Current executable software version}

\begin{table}
\caption{Software metadata (mandatory)}
\label{tab:sw_metadata}
\begin{tabular}{|l|p{5.5cm}|p{5.5cm}|}
\hline
\textbf{Nr.} & \textbf{Software metadata description} & \textbf{Please fill in this column} \\
\hline
S1 & Current software version & v1.0.1 \\
\hline
S2 & Permanent link to executables of this version &
  \url{https://pypi.org/project/buildocc/} \\
\hline
S3 & Legal software license & Apache License 2.0 \\
\hline
S4 & Computing platforms / OS & Linux, macOS, Windows (Python 3.11+) \\
\hline
S5 & Installation requirements \& dependencies &
  \texttt{pip install buildocc}; an LLM provider API key (Anthropic, OpenAI,
  Google) or a local Ollama model \\
\hline
S6 & Link to user manual &
  \url{https://github.com/humanbuildingsynergy/BuildOcc/blob/main/README.md} \\
\hline
S7 & Support email for questions & wooyoung@arizona.edu \\
\hline
\end{tabular}
\end{table}

\credit{\textbf{Wooyoung Jung}: Conceptualization, Methodology, Software, Validation,
Formal analysis, Writing -- original draft, Writing -- review \& editing.}
\competing{The author declares that there are no competing interests.}
\acknowledgments{This material is based upon work partially supported by the National Science Foundation under Grant No.\ 2519054. Any opinions, findings, and conclusions or recommendations expressed in this material are those of the author and do not necessarily reflect the views of the National Science Foundation.}

\appendix
\renewcommand{\thesection}{\Alph{section}}
\setcounter{table}{0}
\renewcommand{\thetable}{\thesection.\arabic{table}}
\makeatletter\@addtoreset{table}{section}\makeatother

\section{ATUS Activity Probability Computation}
\label{app:atus}

This appendix provides the formal description of the activity probability
computation summarized in Section~\ref{sec:software}.

\paragraph{Episode-overlap framing}
ATUS diaries record each respondent's day as a sequence of episodes, each
with a start time $t_s$ (minutes since midnight), stop time $t_e$, and a
six-digit activity code in ATUS.  For each hour $h$, the method identifies which episode was in progress at the
half-hour mark:

\begin{equation}
  c(r,h) = \text{activity code of respondent } r
  \text{ such that } t_s \leq h{:}30 < t_e
  \label{eq:inprogress}
\end{equation}

\noindent
where $r$ indexes individual diary respondents. This in-progress query
avoids the start-time bias that would overrepresent short activities
(under a start-time count, a 5-minute meal and an 8-hour sleep would each count once).

ATUS diaries run from 04:00 to 04:00 the following day, and episode times are recorded
as ordinary clock times. Episodes are therefore placed on a 04:00-anchored axis before
the hourly query, so that an episode crossing midnight is credited to the diary day it
belongs to.

\paragraph{Category probability}
Nine activity categories are defined (Table~\ref{tab:atus_categories}).
The probability of category $k$ at hour $h$, for demographic stratum $s$ and day type
$d \in \{\text{weekday, weekend}\}$, is:

\begin{table}[!ht]
\caption{The nine broad activity categories used in the two-step ATUS sampling
         procedure. ATUS activity codes are six-digit and hierarchical, so a filter
         is a leading prefix of the code; the third column counts the six-digit codes
         mapped to each category.}
\label{tab:atus_categories}
\centering
\begin{tabular}{lp{0.22\linewidth}rp{0.45\linewidth}}
\toprule
\textbf{Category} & \textbf{ATUS code filter} & \textbf{Codes} & \textbf{Representative subcategories} \\
\midrule
Sleeping         & prefix 0101          &   3 & Sleeping; Sleeplessness \\
Work             & prefix 05            &  14 & Work (main job); Work (other job); Security procedures related to work \\
Food preparation & prefix 0202          &   4 & Food and drink preparation; Food presentation; Kitchen and food clean-up \\
Laundry          & code 020102 exactly  &   1 & Laundry \\
Television       & codes 120303, 120304 &   2 & Television and movies; Television (religious) \\
Eating           & prefix 1101          &   2 & Eating and drinking \\
Exercise         & prefix 13            &  42 & Aerobics; Basketball; Biking; Swimming; Weightlifting \\
Travel           & prefix 18            &  31 & Travel related to work; to housework; to personal care \\
Other            & residual             & 201 & Grooming; Reading; Computer use; Social communication \\
\bottomrule
\end{tabular}
\end{table}

\begin{equation}
  P(k \mid s, d, h)
  = \frac{\displaystyle\sum_{r \in R(s,d)} w_r \cdot
          \mathbf{1}[c(r,\, h{:}30) \in k]}
         {\displaystyle\sum_{r \in R(s,d)} w_r}
  \label{eq:catprob}
\end{equation}

\noindent
where $R(s,d)$ is the set of diary respondents matching stratum $s$ on day
type $d$; $w_r$ is the ATUS person-level survey weight (provided by the Bureau of Labor Statistics (BLS) to correct for
sampling design); $\mathbf{1}[\cdot]$ is the indicator function
that equals 1 when respondent $r$ was engaged in a category-$k$ activity at
the half-hour mark $h{:}30$ of hour $h$; and the denominator normalizes the
weighted counts to probabilities.  Evaluating Equation~(\ref{eq:catprob})
for all $h \in \{0,\ldots,23\}$ and all nine categories yields a
$24 \times 9$ weighted probability table per stratum per day type,
which is bundled with the installed package.

\paragraph{Code-level sampling}
Once a category $k$ is sampled from $P(k \mid s, d, h)$, a specific activity
code is drawn from within that category proportional to the total diary minutes
logged across all stratum respondents for each code. This gives common activities
realistic relative frequencies (e.g., sleeping in a bed is drawn far more often
than napping on a sofa, even though both map to the ``sleeping'' category).

\paragraph{Work-from-home probability}
Finally, the WFH probability $p_\text{WFH}$ for each stratum is estimated
from the ATUS \texttt{TEWHERE} variable --- the ``where were you when you did
this activity?'' field --- restricted to employed respondents who worked on
their diary day.  $p_\text{WFH}$ is the share of the stratum's survey-weighted work minutes reported at
home (the at-home column of Table~\ref{tab:strata_rationale}).  A per-day Bernoulli draw from $p_\text{WFH}$
then determines whether the simulated agent works from home on a given day.

\section{Memory Retrieval and Reflection Mechanics}
\label{app:memory}

This appendix provides the mathematical detail for the memory retrieval and
reflection logic implemented in \texttt{MemoryStream}, following the generative
agent architecture of Park et al.~\cite{park2023}.

\paragraph{Retrieval scoring}
At each simulation timestep $t$, every stored memory entry $m$ receives a
combined retrieval score $s(m,t) \in [0,1]$ that balances how recent the
entry is against how important it was judged to be when written:

\begin{equation}
  s(m,\,t) = 0.5\;\tilde{r}(m,t) + 0.5\;\tilde{i}(m)
  \label{eq:retrieval}
\end{equation}

\noindent
where $\tilde{r}(m,t)$ is the normalized recency score and $\tilde{i}(m)$ is
the normalized importance score, both in $[0,1]$ and weighted equally.

The recency component decays exponentially with a 24-hour half-life:

\begin{equation}
  \tilde{r}(m,t) = 2^{-(t-t_m)/24\,\text{h}}
  \label{eq:recency}
\end{equation}

\noindent
where $t_m$ is the simulation time (in hours) at which entry $m$ was written,
and $(t - t_m)$ is the elapsed time in hours since writing.
The score equals 1.0 at the moment of creation and halves every 24 simulation
hours, so memories from the previous day retain half their recency
weight while week-old entries contribute near zero.

The importance component normalizes the integer score $i(m)$ assigned by the
LLM at write time:

\begin{equation}
  \tilde{i}(m) = \frac{i(m)}{10}
  \label{eq:importance}
\end{equation}

\noindent
where $i(m) \in \{1, \ldots, 10\}$ is an integer on a scale from 1 (routine,
nothing unusual) to 10 (rare, extreme situation), self-assigned by the
agent's LLM in the same JSON response that produces the behavioral action.
The top-$k$ entries ranked by $s(m,t)$ are injected into the next prompt
($k = 5$ by default).

\paragraph{Reflection trigger}
An importance accumulator $A$ tracks the cumulative importance of all entries
written since the last reflection.  After each entry $m$ is stored,
$A \leftarrow A + i(m)$.  When $A \geq \theta$ (default $\theta = 100$),
reflection is triggered: a dedicated LLM call receives the 30 most recent
entries by simulation time and returns exactly three synthesized behavioral
insights (e.g., ``this occupant pre-cools before peak hours to avoid peak
tariffs'').  Each insight is written back into the stream as a new entry with
importance~9.0, and $A$ resets to zero.
At average importance 7--10 (high-salience events), reflection fires every 10--14 timesteps ($\approx$2.5--3.5 simulation hours); at average
importance 2--3 (routine observations), every 33--50 timesteps ($\approx$8--12 simulation hours). In
the measured runs of Section~\ref{sec:software}, reflection fired 1.5 times per
simulated day.

\section{Demographic Strata Selection Rationale}
\label{app:strata}

The four strata were selected to maximize variation in the two behavioral
dimensions most consequential for residential demand response research:
\emph{peak-hour occupancy schedule} and \emph{household composition}.
The selection is grounded in the ATUS 2022--2023 microdata that drives the
platform itself~\cite{mitra2020,koupaei2022,vosoughkhosravi2023}.

Table~\ref{tab:strata_rationale} summarizes the occupancy metrics and sample
sizes derived from ATUS 2022--2023 ($n = 16{,}684$ respondents).  O1 and O3
are absent from the home for the majority of daytime hours (09:00--17:00),
while O2 and O4 are home for most of the day: 60.9\% and 62.6\% of O1 and O3 weekday
hours in that window are spent in work activity, against 0.4\% and 2.4\% for O2
and O4.  Within each occupancy cluster, household composition
is varied to capture differences in thermostat authority and social dynamics:
O1 lives alone, O2 is a couple, O3 has children at
home, and O4 carries no household restriction~\cite{gyamfi2011price,albadi2008summary}.  O4 has the smallest sample
($n = 677$).
Additional strata (multi-adult non-family, student dormitories) would draw from the
remaining 10,073 ``other'' respondents but would fragment already thin subsamples
without a clearly defined demand response participation profile.

\begin{table}[!ht]
\caption{ATUS 2022--2023 occupancy metrics and sample sizes by stratum.
         \% working 09--17 is the mean weighted probability of being engaged in work
         activity during the 09:00--17:00 window on weekdays (Appendix~\ref{app:atus}).
         Location columns split each stratum's work time (ATUS activity codes 05,
         minutes weighted by \texttt{TUFINLWGT}, episodes with a valid location code)
         across the respondent's home, a workplace, and elsewhere; the three sum to
         100\%. For O2 and O4 the split rests on the small amount of work time those
         strata report. The at-home share is the work-from-home probability of
         Section~\ref{sec:software}.}
\label{tab:strata_rationale}
\centering
\begin{tabular}{llrrrrr}
\toprule
\textbf{ID} & \textbf{Stratum} & \textbf{\% working} & \textbf{\% at} & \textbf{\% at} & \textbf{\% else-} & \textbf{$n$} \\
            &                  & \textbf{09--17 (wkdy)} & \textbf{workplace} & \textbf{home} & \textbf{where} & \\
\midrule
O1 & Employed single adult & 60.9 & 70.2 & 24.3 &  5.6 & 1,393 \\
O2 & Retired couple        &  0.4 &  8.1 & 50.1 & 41.8 & 2,351 \\
O3 & Employed parent       & 62.6 & 68.9 & 25.7 &  5.4 & 2,190 \\
O4 & Not-employed adult    &  2.4 & 26.6 & 68.7 &  4.6 &   677 \\
\bottomrule
\end{tabular}
\end{table}

\end{document}